# Broadband Terahertz Modulators based on MEMS-Reconfigurable Mesh Filters

M. Unlu[1], M. R. Hashemi[1], C. W. Berry[1], S. Li[1], S.-H. Yang[1], and M. Jarrahi[1]*

[1]Electrical Engineering and Computer Science Department, University of Michigan, Ann Arbor, MI 48109, United States

*Correspondence to: mjarrahi@umich.edu

**Abstract:** Active tuning and switching of electromagnetic properties of materials is of great importance for controlling their interaction with electromagnetic waves. Superconductors are the only natural materials that exhibit diamagnetic switching at their critical temperatures. Here, we demonstrate a new class of meta-surfaces with electrically-induced diamagnetic switching capability at room temperature. Structural configuration of the subwavelength meta-molecules determines their magnetic response to an external electromagnetic radiation. By reconfiguration of the meta-molecule structure, the strength of the induced magnetic field in the opposite direction to the external magnetic field is varied and the diamagnetic state of the meta-surface is altered, consequently. We demonstrate a significant change in the relative permeability and permittivity of a custom-designed meta-surface with diamagnetic switching capability at terahertz frequencies, enabling terahertz intensity modulation with record high modulation depths and modulation bandwidths through a fully integrated device platform at room temperature.

Reconfigurable metamaterials, artificial electromagnetic media with tunable and switchable electromagnetic characteristics, have attracted extensive attention due to their unique capabilities for routing and manipulating electromagnetic waves. Tuning the electromagnetic characteristics of metamaterials can be achieved by controlling the structure and arrangement of their so called *meta-molecule* building blocks. In this regard, a number of tunable metamaterials based on phase-change materials (*1-3*), semiconductors (*4-6*), graphene (*7-9*), superconductors (*10-12*), and electromechanical structures (*13-15*) activated by optical, electrical, magnetic, and thermal stimuli have been demonstrated. Consequently, new paradigms for switching electromagnetic waves and tuning electromagnetic phase, polarization, propagation direction, and beam shape have emerged, which would not have been possible by use of naturally existing substances (*16*). In spite of the significant progress in the field, the demonstrated tunable metamaterials are often limited in their operation bandwidth, mainly due to the resonant nature of their meta-molecule constituents. To attain the flexible design space that transformation optics demands, new artificial materials with extraordinary permittivity and permeability tunability over a broad range of frequencies are required.

Extreme diamagnetic switching over a broad range of frequencies is a property that has been only offered by superconductors when cooled below their critical temperature (around liquid helium temperature for most superconductors). The so called Meissner effect (*17*) repels the magnetic field from the bulk of the superconductor, excluding the field lines completely from the superconductor region. Here we demonstrate broadband, extreme diamagnetic switching at room temperature, for the first time, through a new class of electrically-reconfigurable meta-surfaces. By reshaping the structural configuration of the strongly coupled meta-molecule unit cells, their collective magnetic response to an incident electromagnetic wave is altered. Accordingly, efficient electromagnetic flux transmission through the meta-surface and strong electromagnetic exclusion from the meta-surface region is observed in the weak and strong diamagnetic states, respectively (Fig. 1). Dynamic switching of the meta-surface diamagnetic state over a broad range of electromagnetic frequencies will offer new opportunities for signal handling, electromagnetic switching, beam steering, and cloaking. We demonstrate a fully integrated platform for modulating the intensity of terahertz waves interacting with a custom-designed meta-surface with switchable diamagnetic characteristics, enabling terahertz intensity modulation with record high modulation depths (> 70%) and modulation bandwidths (> 1.5 THz) at room temperature. We have deliberately designed our proof-of-concept meta-surface prototype for operation at terahertz frequencies, since existing modulation schemes in the visible and infrared regime based on carrier injection/depletion in solid-state devices (*18*, *19*), Mach-Zehnder interferometers (*20*), Fabry-Perot filters (*21*), liquid crystals (*22*), magneto-optic effects (*23*), deformable mirrors (*24*), and beam deflectors (*25*, *26*), have difficulty in offering high-performance modulation specifications at terahertz frequencies.

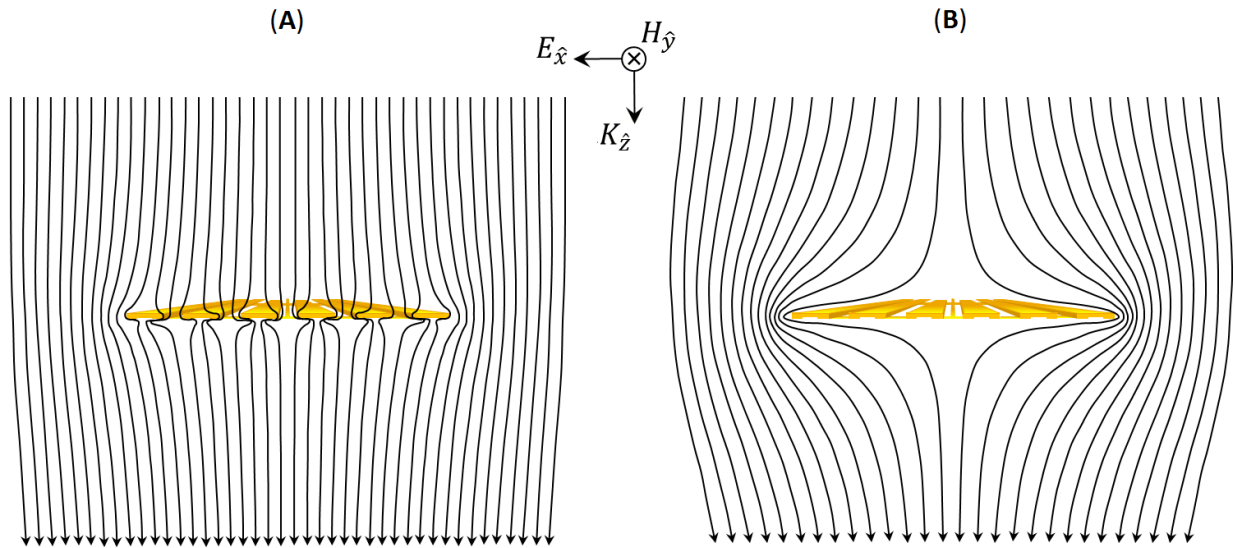

**Fig. 1.** Electromagnetic interaction with the meta-surface with switchable diamagnetism. Electromagnetic flux penetrates through the meta-surface during the weak diamagnetic state (**A**) and is forced to curve away and exclude the meta-surface region during the strong diamagnetic state (**B**).

The schematic diagram and operation principle of the meta-surface with diamagnetic switching capability is shown in Fig. 2. It consists of an array of vertically-oriented (along the *y*-axis) Au membranes, suspended above a silicon substrate. The anchors of the suspended membranes are placed on a $SiO_2$ layer to electrically isolate them from the underlying silicon substrate. When a voltage difference is applied between the Au membranes and the silicon substrate, the induced electrostatic force deflects the Au membranes, moving them into contact with an array of horizontally-oriented (along the *x*-axis) Au patches on the SiO$_2$ layer (*27*). All the Au membranes and patches are connected through vertically-oriented Au traces such that the entire array of suspended membranes can be deflected simultaneously by controlling the voltage difference between the meta-surface and the underlying silicon substrate.

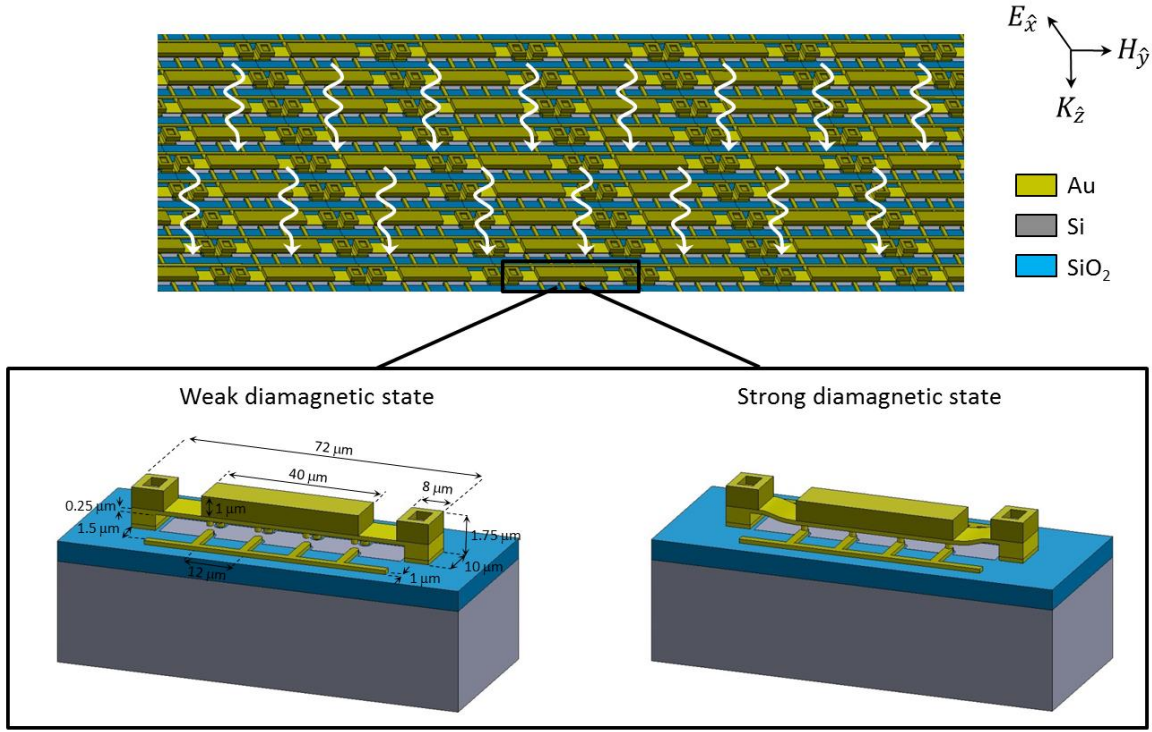

**Fig. 2:** The operation principle of the meta-surface with diamagnetic switching capability. (**A**) The schematic of the meta-surface, which consists of an array of vertically-oriented Au membranes suspended above a silicon substrate. Depending on the voltage difference between the Au membranes and the silicon substrate, the Au membranes can be suspended above the silicon substrate (the weak diamagnetic state) or be in contact with an array of horizontally-oriented Au patches on the substrate (the strong diamagnetic state). A relatively thick metal layer is used in the center of the Au membranes to assist with the flatness of the contact areas and achieve high spring constants required for high speed switching. The gap between the Au membranes and the substrate (0.5 µm), and the contact dimple height (0.25 µm) are chosen to achieve low switching voltages (30 V) and high switching speeds (> 20 KHz) while accounting for possible bending in the Au moving membranes as a result of uncompensated stress of the Au membranes.

When an electromagnetic wave impinges on the meta-surface, the free electrons in the meta-surface are accelerated by the applied Lorentz force, generating a magnetic field opposing the incident magnetic field. The magnitude of the induced magnetic field determines the diamagnetic response of the structure. Before deflecting the Au membranes, the strength of the induced magnetic field in response to a horizontally-polarized incident electromagnetic wave is weak for subwavelength meta-molecule feature sizes in the horizontal direction. This weak diamagnetic behavior is the result of the metal discontinuities, which prevent the electrons from accelerating in the horizontal direction to gain kinetic energy and produce a considerable opposing magnetic field. After deflecting the Au membranes, allowing them to contact the Au patches, the metal electrons can accelerate and gain kinetic energy as a result of the induced

Lorentz force. In this state, the strength of the induced magnetic field in response to the horizontally-polarized incident electromagnetic wave is strong for subwavelength meta-molecule feature sizes in the vertical direction. This strong diamagnetic behavior is the result of efficient energy transfer from the incident electromagnetic wave to the kinetic energy of the metal electrons, producing an opposing magnetic field that cancels the incident field in the meta-surface region.

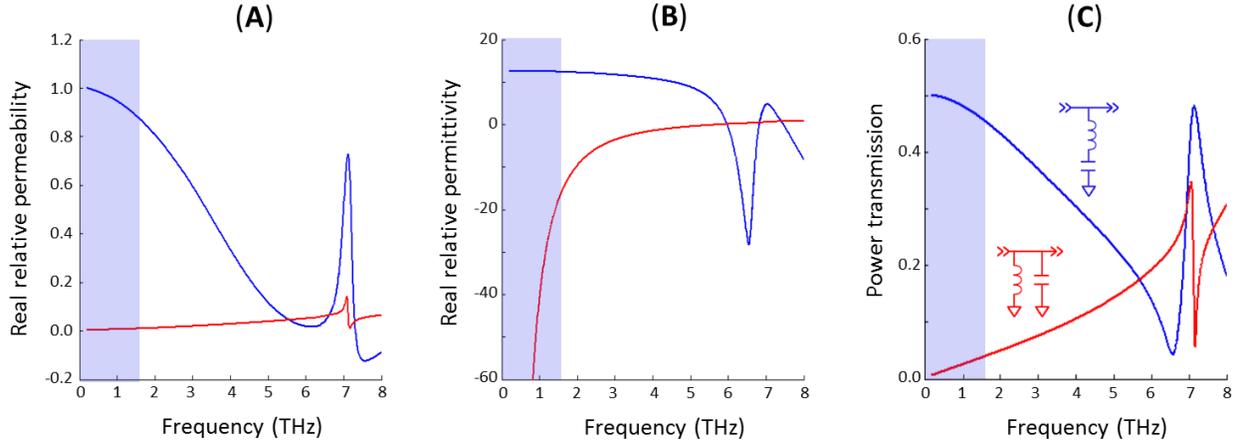

**Fig. 3:** The extracted relative permeability, relative permittivity, and electromagnetic power transmission through the designed meta-surface for a horizontally-polarized incident electromagnetic wave during the weak diamagnetic state (blue curves) and strong diamagnetic state (red curves) are illustrated in (**B**), (**C**), and (**D**), respectively, exhibiting an extraordinary switching of the scattering parameters over a broad frequency band (highlighted in blue). Electromagnetic wave interaction with the designed reconfigurable meta-surface can be also explained by the meta-surface equivalent circuit model (inset), where the horizontally-oriented metal lines function as inductors and the horizontal gaps between metal stripes function as capacitors.

A unique advantage of the presented reconfigurable meta-surface is that the strong and weak diamagnetism are exhibited over a frequency band set by the meta-surface feature size in the vertical and horizontal direction, respectively. Therefore, it can offer broadband diamagnetic switching through structural miniaturization. By appropriate choice of the vertical distance between the Au patches (12 μm) and the horizontal distance between the Au membranes (16 μm), we have designed a reconfigurable meta-surface with broadband diamagnetic switching capability over a 1.5 THz frequency band. Using a finite-element-based full-wave electromagnetic solver (ANSYS HFSS), we have analyzed the interaction of a horizontally-polarized electromagnetic wave with the designed meta-surface and extracted the effective permeability and permittivity from the calculated scattering parameters using a robust material constitutive effective parameters retrieval method (*28*, *29*). The extracted permeability of the designed meta-surface indicates extreme diamagnetic switching (from $\mu_r > 0.9$ to $\mu_r < 7\times10^{-3}$) over a 1.5 THz frequency band (Fig. 3A), a functionality that has been only offered by superconductors when cooled below their critical temperatures.

Similar to superconductors, the exhibited diamagnetic switching of the designed reconfigurable meta-surface is accompanied by a significant change in permittivity over a broad frequency band (Fig. 3B). This is because the strength of the induced electric field opposing the external electric field is directly affected by the magnitude of the electric current induced on the meta-surface. Therefore, the negligible opposing electric field in the weak diamagnetic state allows the incident electric field to penetrate through the meta-surface structure, exhibiting a positive real permittivity, set by the substrate permittivity. Similarly, the opposing electric field in the strong diamagnetic state cancels the incident electric field in the meta-surface region, exhibiting a negative real permittivity, similar to a solid slab of metal. As a result, an efficient electromagnetic flux transmission through the meta-surface and a strong electromagnetic flux exclusion from the meta-surface region are exhibited in the weak and strong diamagnetic states, respectively, for both the electric and magnetic flux.

Figure 3C shows the estimated electromagnetic power transmission through the designed reconfigurable meta-surface for a horizontally-polarized incident electromagnetic wave. It shows the capability of the designed reconfigurable meta-surface for modulating the intensity of terahertz waves with more than 90% modulation depth, over a 1.5 THz frequency band. This unprecedented modulation performance is made possible by the broadband permeability/permittivity switching capability of the designed reconfigurable meta-surface and can be extended even further to offer higher modulation depths and modulation bandwidths through further structural miniaturization. The 50% electromagnetic flux transmission in the weak diamagnetic state is bound by the Fresnel reflections at the silicon-air interfaces of the designed reconfigurable meta-surface mounted on a silicon lens. The 3 dB signal loss associated with Fresnel reflections at silicon-air interfaces can be eliminated by removing or thinning the silicon substrate under the device active area. It should be mentioned that strong diamagnetism is achieved not only through complete physical contact between the Au membranes and Au patches, but also exhibited for nanometer-scale metal separations. For nanometer-scale separations, the large capacitance between the Au membranes and Au patches shunts the two metal layers at high frequencies (Fig. 4). This allows diamagnetic switching at relatively low voltages and enables reliable device operation by eliminating the need for applying high electrostatic forces and direct metal-to-metal contact (*30*).

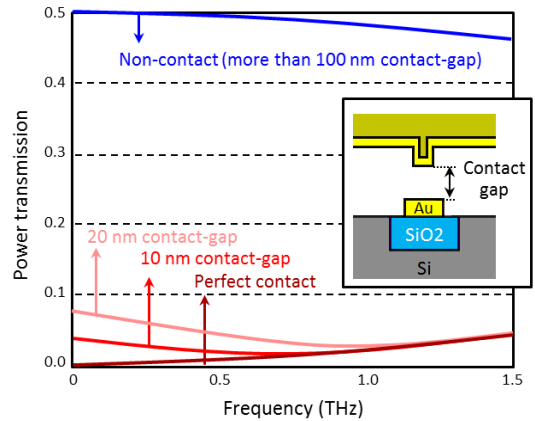

**Fig. 4:** Estimated power transmission of a horizontally-polarized electromagnetic wave through the designed meta-surface as a function of the gap between the Au moving membrane and the Au patches. While efficient transmission of the horizontally-polarized electromagnetic waves is maintained for contact gaps as small as 100 nm, modulation depth levels of more than 90% are expected even for contact gaps as large as 10 nm.

The extraordinary switching of the scattering parameters of the presented meta-surface has been utilized to develop a room-temperature terahertz intensity modulator with unprecedented modulation depth and bandwidth. The modulator prototype is fabricated on a high-resistivity silicon substrate. The detailed description of the fabrication process is illustrated in Fig. 5.

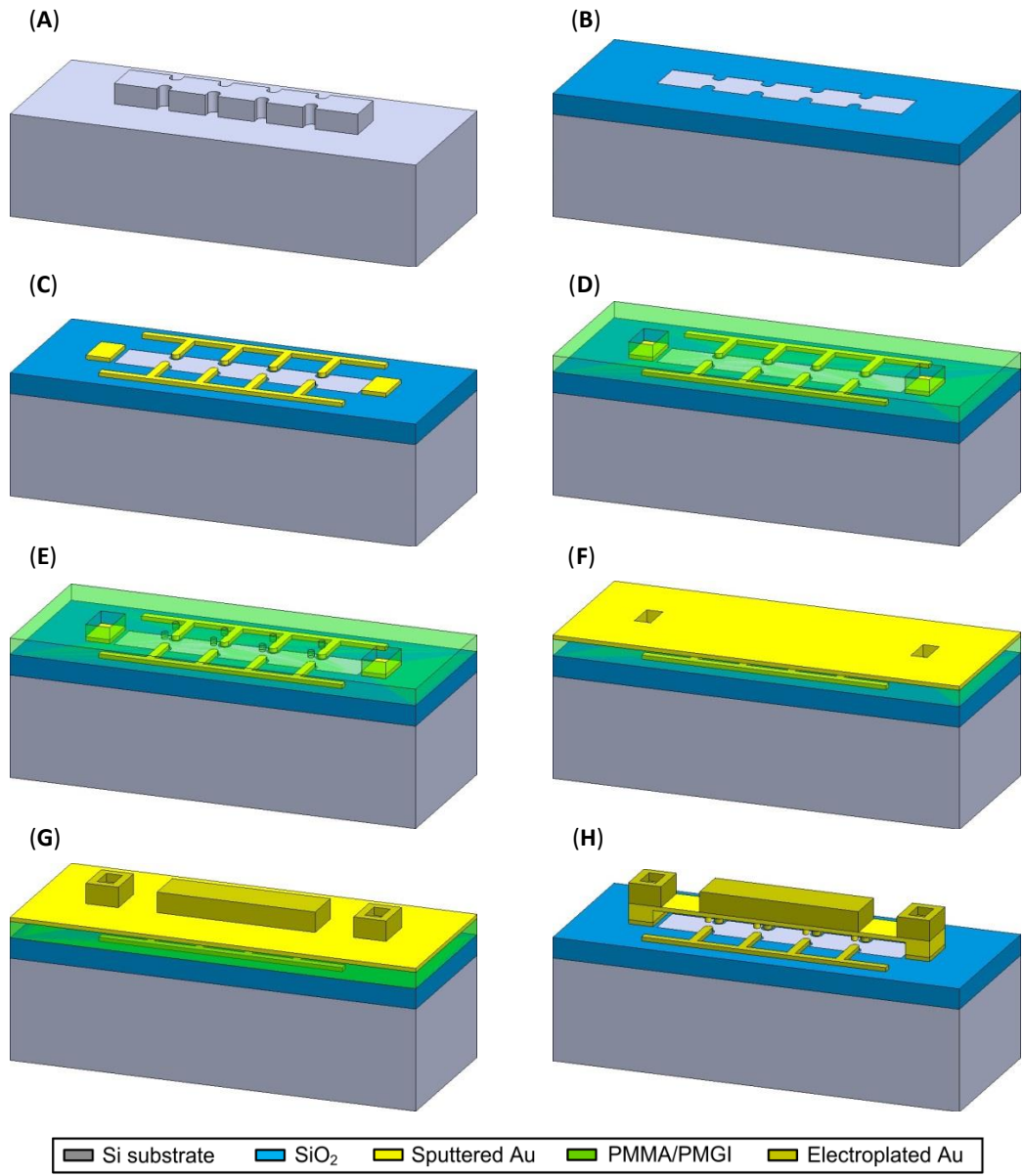

**Fig. 5:** Fabrication process. The fabrication process starts with defining the $SiO_2$ isolation layer areas by Si reactive ion etching (**A**), followed by $SiO_2$ deposition, using plasma enhanced chemical vapor deposition. Subsequently, the wafer is planarized, using chemical mechanical polishing (**B**). The horizontally-oriented Au patches and bias lines are then formed by sputtering Ti/Au/Ti (100/1000/100 Å) followed by lift-off (**C**). A PMMA/PMGI sacrificial layer (0.5 μm) is then spin coated and patterned for the anchor areas (**D**) and contact dimples (**E**) using two separate masks. Next, a Ti/Au layer (100/2500 Å) is deposited using sputtering as the seed layer for electroplating (**F**). Subsequently, a 1 μm-thick Au layer is defined by optical photolithography and selectively electroplated for the anchor areas and the thick metal section in the center of the Au moving membranes (**G**). Finally, the seed layer and the sacrificial layer are removed using wet etching, and the modulator is released using critical point drying (**H**).

Figure 6A shows the schematic diagram and the scanning electron microscope (SEM) image of a fabricated terahertz modulator prototype. The meta-molecule elements are electrically connected such that the entire meta-surface structure can be switched between the weak diamagnetic state (modulation 'OFF' mode) and the strong diamagnetic state (modulation 'ON' mode), by controlling the voltage difference between the meta-surface and the underlying substrate, $V_{Switch}$. To simplify the terahertz modulator characterization, a square metallic aperture of ~ 1 mm × 1 mm is fabricated around the meta-surface region to ensure that the transmitted terahertz power through the structure has fully interacted with the device active area.

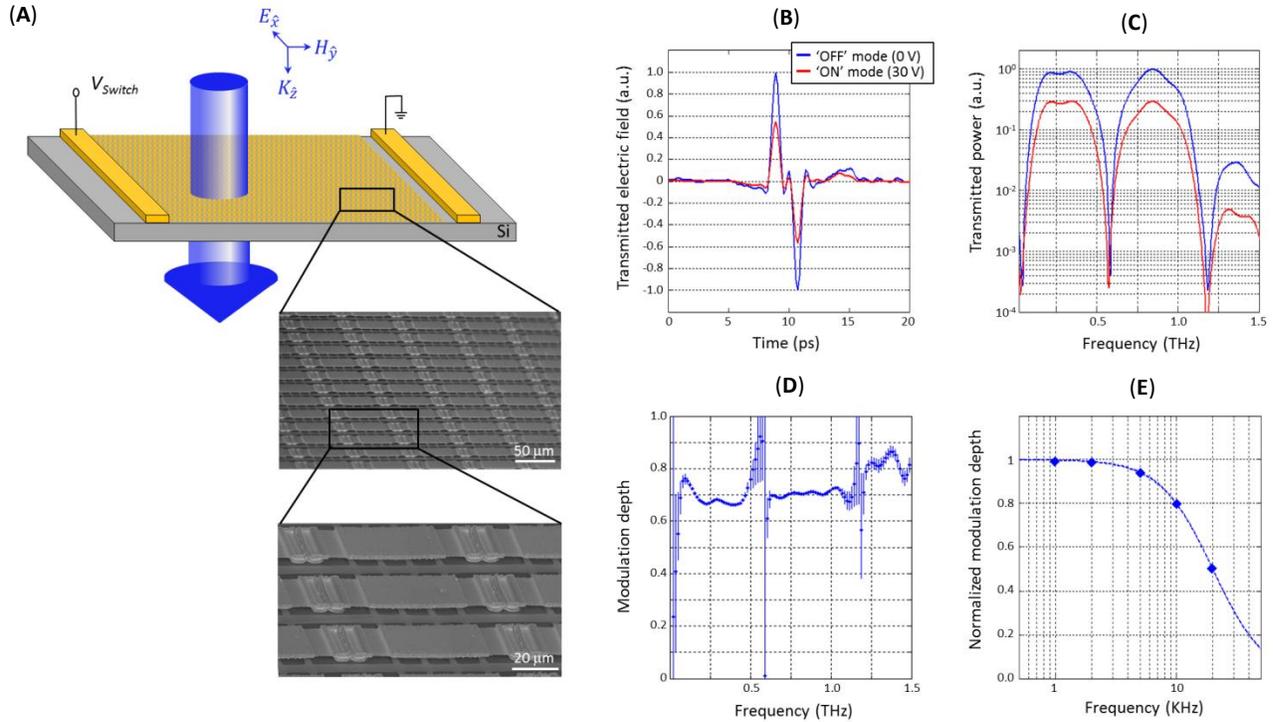

**Fig. 6:** Performance of the proof-of-concept terahertz modulator based on the designed meta-surface with diamagnetic switching capability. (**A**) The schematic diagram and the SEM image of the fabricated terahertz modulator prototype. (**B**) Electric field of a transmitted terahertz pulse through the terahertz modulator prototype at a bias voltage of 0 V (modulation OFF mode) and 30 V (modulation ON mode). (**C**) The calculated power transmission spectrum of the terahertz pulses incident on the modulator prototype over the 1.5 THz frequency range during the modulation OFF and ON modes. The observed spectral dips are the result of the apertures used for focusing terahertz pulses onto the device under test. (**D**) Modulation depth of the modulator prototype, calculated using the power transmission spectra during the modulation OFF and ON modes, indicating more than 70% modulation depth over the 1.5 THz frequency band. The data error bars are calculated using the noise power level of the time-domain terahertz spectroscopy setup used for characterizing the modulator performance. The calculated modulation depth is quite accurate for the 1.5 THz frequency range, except at frequencies that the apertures used for focusing terahertz pulses onto the modulator prototype severely attenuate power transmission. (**E**) The dynamic characteristics of the terahertz modulator prototype, characterized by measuring the electric field of the transmitted terahertz pulses through the modulator prototype while alternating the bias voltage between 0 V and 30 V. The calculated terahertz modulation depth as a function of the modulation speed indicates modulation speeds exceeding 20 KHz.

The performance of the fabricated terahertz modulator is characterized in a time-domain terahertz spectroscopy setup. By measuring the transmitted electric field of a horizontally-polarized terahertz pulse through the implemented terahertz modulator (Fig. 6B), the spectrum of the transmitted power during the modulation OFF mode ($V_{Switch} = 0$ V) and modulation ON mode ($V_{Switch} = 30$ V) is calculated (Fig. 6C). The observed spectral dips are the result of the apertures used for focusing terahertz pulses onto the device under test. The modulation depth of the fabricated modulator prototype is calculated using the power transmission spectra during the modulation OFF and modulation ON modes, indicating more than 70% modulation depth over the 1.5 THz frequency band (Fig. 6D). The data error bars in Fig. 6D are calculated using the noise power level of the time-domain terahertz spectroscopy setup used for characterizing the performance of the modulator prototypes. As illustrated in Fig. 6D, the calculated modulation depth is quite accurate over the 1.5 THz frequency range, except at the frequencies that the apertures used for focusing the terahertz pulses onto the modulator active area severely attenuate the power transmission. By comparing the power transmission spectrum of the terahertz pulses incident on the modulator prototype with the power transmission spectrum of the terahertz pulses incident on the silicon substrate through the same aperture, the signal attenuation of the modulator prototype is calculated, indicating a signal attenuation of less than 3.3 dB for the 1.5 THz frequency band. Finally, the dynamic characteristics of the fabricated terahertz modulator is analyzed by measuring the electric field of the transmitted terahertz pulses through the modulator while alternating the applied voltage between 0 V and 30 V. Figure 6E shows the calculated terahertz modulation depth as a function of the modulation speed, indicating modulation speeds exceeding 20 KHz.

It should be noted that the achieved modulation depth from the fabricated modulator prototype is ~20% lower than the theoretically estimated modulation depth of more than 90% in the 1.5 THz frequency range. We attribute this mainly to fabrication misalignments, which result in a non-uniform spacing between the Au membrane and Au patches of meta-molecule elements. This prevents the required contact between the Au membrane and Au patches for a portion of the meta-molecules during the strong diamagnetic state. This hypothesis is supported by the observed increase in the calculated modulation depths at higher frequencies, reaching 85% modulation depth in the 1-1.5 THz frequency range. In spite of the capacity of the presented modulator to offer higher modulation depths through optimized fabrication processes, the achieved modulation depth is the highest reported among previously demonstrated terahertz intensity modulators, in general (*31*), and 5 times higher than the demonstrated broadband terahertz modulators with similar modulation voltages and modulation speeds, specifically (*32*). This unprecedented modulation performance is made possible by the significant change in the relative permeability and permittivity of the implemented meta-surface with diamagnetic switching capability, over a broad frequency band. Moreover, the demonstrated broadband diamagnetic switching capability, which has been only offered by superconductors at their critical temperatures, is achieved in a fully integrated, electrically controlled, room temperature platform. The diamagnetic switching principles of the presented meta-surface are universal and can be extended to optical and infrared frequency ranges by further structural miniaturization. Additionally, the diamagnetic switching speed of the proof-of-concept reconfigurable meta-surface, which has been limited by the restoring force of the designed Au membranes, can be further increased to gigahertz-range switching speeds by structural miniaturization or use of higher spring constant membranes, like graphene (*33*).

**References:**

1. R. E. Simpson *et al., Nat. Nanotech.* **6**, 501 (2011).
2. T. Driscoll *et al*., *Science* **325,** 1518 (2009).
3. M. Seo *et al., Nano Lett.* **10**, 2064 (2010).
4. H. T. Chen *et al., Nature* **444,** 597 (2006).
5. S. Zhang *et al., Nat. Commun*. **3,** 942 (2012).
6. H. T. Chen *et al., Nat. photon*. **3**. 148 (2009).
7. H. G. Yan *et al., Nat. Nanotech*. **7,** 330 (2012).
8. L. Ju *et al., Nat. Nanotech*. **6,** 630 (2011).
9. S. H. Lee *et al., Nat. Mater.* **11**, 936 (2012).
10. F. Gomory *et al., Science* **335,** 1466 (2012).
11. M. Ricci, N. Orloff, S. M. Anlage, *Appl. Phys. Lett*. **87,** 034102 (2005).
12. B. Wood *et al., Nat. Mater*. **7,** 295 (2008).
13. I. M. Pryce, K. Aydin, Y. A. Kelaita, R. M. Briggs, H. A. Atwater, *Nano Lett.* 10, 4222 (2010).
14. C. W. Berry, J. Moore, M. Jarrahi, *Opt. Express* **19**, 1236 (2011).
15. A. Q. Liu, W. M. Zhu, D. P. Tsai, N. I. Zheludev, *J. Opt.* 14, 114009 (2012).
16. N. I. Zheludev, Y. S. Kivshar, *Nat. Mater*. **11**, 917 (2012).
17. I. F. London, *Superfluids: Macroscopic Theory of Superconductivity* 1, 2nd ed., Dover (1961).
18. A. Liu *et al*., *Nature* **427**, 615 (2004).
19. M. Liu *et al*., *Nature* **474**, 64 (2011).
20. M. Jarrahi, T. H. Lee, D. A. B. Miller, *Photon. Technol. Lett.* **20**, 517 (2008).
21. G. Lammel, S. Schweizer, S. Schiesser, P. Renaud, *J. Microelectromechanical Systems* **11**, 815 (2002).
22. A. M. Wiener, D. E. Leaird, J. S. Patel, J. R. Wullert, *Journal Quantum Electron.* **28**, 908 (1992).
23. W. E. Ross, D. Psaltis, R. H. Anderson, *Opt. Eng.* **22**, 485 (1983).
24. O. Solgaard, F. S. A. Sandejas, D. M. Bloom, *Opt. Lett.* **17**, 688 (1992).
25. M. Jarrahi, R. F. W. Pease, D. A. B. Miller, T. H. Lee, *Appl. Phys. Lett.* **92**, 014106 (2008).
26. M. Jarrahi, R. F. W. Pease, D. A. B. Miller, T. H. Lee, *J. Vacuum Sci.& Technol. B*. **26**, 2124 (2008).
27. J. B. Muldavin, G. M. Rebeiz, *IEEE Microwave Wireless Comp. Lett*. **11**. 373 (2001).
28. X. Chen *et al., Phys. Rev. E.* **70**, 016608 (2004).
29. D. R. Smith, D. C. Vier, T. Koschny, C. M. Soukoulis, *Phys. Rev. E.* **71**, 036617 (2005).
30. R. Maboudian, R. T. Howe, *J. Vac. Sci. Technol. B*. **15**, 1 (1997).
31. M. Rahm, J. S. Li, W. J. Padilla, *J. Infrared Millimeter & Terahertz Waves* **34**, 1 (2012).
32. B. Sensale-Rodriguez *et al., Nat. Commun.* **3**, 780 (2012).
33. I. W. Frank, D. M. Tanenbaum, A. M. van der Zande, P. L. McEuen, *J. Vac. Sci. Technol. B*. **25**, 2558 (2007).

**Acknowledgments:** Jarrahi's group at the University of Michigan would like to thank the Lurie Nanofabrication Facility staff members for their help with device fabrication and gratefully acknowledge the financial support from National Science Foundation Sensor and Sensing Systems division (Contract # 1030270) and Army Research Office Young Investigator Award (Contract # W911NF-12-1-0253).

**Author Contributions:** Mehmet Unlu fabricated the device prototypes. Mohammad Reza Hashemi performed all of the electromagnetic analysis. Christopher Berry performed all of the measurements. Shenglin Li assisted with the electromagnetic analysis. Shang-Hua Yang assisted with device fabrication. Mona Jarrahi came up with the idea, supervised the project, and wrote the manuscript. All authors discussed the results and commented on the manuscript.